\documentclass[doublecol]{epl2} 
\usepackage{amsmath}
\usepackage{amsfonts}
\usepackage{mathtools}
\usepackage[svgnames]{xcolor}
\usepackage[colorlinks=true, linkcolor=Maroon, urlcolor=Blue]{hyperref} 
\usepackage{color}
\usepackage{microtype}
\DisableLigatures{encoding = *, family = *}
\usepackage[figuresleft]{rotating}
\usepackage{relsize}
\usepackage{float}
\usepackage{subfigure}
\usepackage{graphicx}
\hypersetup{
  colorlinks,
  citecolor=MidnightBlue,
  linkcolor=DarkRed,
  urlcolor=Blue}

\def\pra{Phys. Rev. A }
\def\prl{Phys. Rev. Lett. }

\title{Playing a true Parrondo's game with a three state coin on a quantum walk}
\shorttitle{True Parrondo's game} 

\author{Jishnu Rajendran\inst{1} \and Colin Benjamin\inst{1}}
\institute{ School of Physical Sciences, National Institute of Science Education \& Research, HBNI, Jatni-752050,\ India                    
  \inst{1} 
}
\abstract{{Playing a Parrondo's game with a qutrit is the subject of this paper. We show that a true quantum Parrondo's game can be played with a 3 state coin(qutrit) in a 1D quantum walk in contrast to the fact that playing a true Parrondo's game with a 2 state coin(qubit) in 1D quantum walk fails in the asymptotic limits.}}

\begin{document}
\maketitle
\section{Introduction}
Quantum walks(QW), motivated from classical random walks, have proven to be of great utility in simulating many physical systems \cite{two_particle,Marquez-Martin2016} as well as in developing better quantum algorithms \cite{algo,qcomp1,qcomp2}, such as the universal computational primitive\cite{universal}. Similar to a classical random walk, quantum walk can be described in a one-dimensional lattice with walker starting from the origin, however, unlike the classical random walk here the walker is described quantum mechanically by a wave function. Similar to a classical random walk, quantum walks consist of a walker and a coin. In case of classical random walks, the walker starts from origin moves left or right according to a coin toss, say if the coin toss yields head the walker moves right otherwise moves left. Probability distribution of the classical random walk in a one-dimensional lattice after $n$ steps is a \textit{binomial distribution} \cite{classical}. In case of quantum walks the walker is a quantum object. The coin corresponding to a quantum walker in case of a two-state(head and tail) coin is a qubit. Similar to the classical random walks, in quantum walks if the coin toss yields head the walker moves towards right otherwise left. In addition to head or tail, the coin in quantum case can be in a superposition of head and tail and then walker will move to a corresponding superposition of left and right lattice sites. The probability distribution of the quantum walk is a two-peaked distribution peaking around the two edges, which is evidently different from that of classical random walk where peak is at origin\cite{Andraca}. Similar to a quantum walk with two state coin(qubit), one can devise a quantum walk with a three-state coin too which is a qutrit. {Quantum walk with a three state coin is not new, in fact studies have been attempted in context of localization\cite{Inui,qutrit1}, weak limit\cite{Falkner} and coin eigenstates\cite{qutrit2} of a one dimensional quantum walk with qutrits.} Qubits are the quantum analogue for head and tail, similarly, qutrit is the quantum analogue of a three-state coin with a head, tail and a side. Similar to the two-state coin quantum walk described above, for a three-state coin quantum walk, if the coin lands on its head the walker moves right, if on tails walker moves to left and if the coin falls on side then the walker stays in the same position, { this wait state of walker staying in the same position is the main difference between qutrit and qubit quantum walks}. In case of classical random walk the walker with three state coin moves toward right when the coin lands on its head, moves left when coin lands on its tail but there is also a small probability that the coin falls on its side, then the coin toss is repeated this is one way of making analogy with the quantum walk with qutrit(three state coin).

The relevance of game theory spreads throughout many scientific fields. Quantum effects introduced to the classical game theory led to the development of quantum game theory, where the probabilities are taken in to account with quantum effects like superposition and interference\cite{q_game}. Parrondo's games\cite{parrondo}, in its simplest form, is a gambling game consisting two games A and B. whose outcome is determined by the toss of a biased coin. When a player plays each of them individually it results in losing whereas if played alternatively can result in a winning outcome. This apparent paradox when each of these games is losing when played individually but when played alternately or in some other deterministic or random sequence (such as $ABB\ldots$,$ABAB\ldots$, etc.) can become a winning game is known as Parrondo's paradox. Originally J.~M.~R.~Parrondo devised this paradox to provide a mechanism for Brownian ratchets where a directed motion can be harnessed from Brownian motion. {In Ref.\cite{maximal_parrondo} Parrondo's games are analysed using classical and quantum Markov Chains. In Ref.~\cite{superactivation} a super-activation-like effect for the capacity of classical as well as quantum communication channels with memory is constructed with the help of Parrondo's paradox.} The previous attempt for a Parrondo's paradox with a single coin qubit on quantum walks failed in asymptotic limits\cite{minli,flitney}. In one of our previous works, we showed how to realize a genuine quantum Parrondo's paradox in the asymptotic limits not with a single coin but with two coins\cite{previous}. In this work, our aim is to replicate the quantum Parrondo's paradox in asymptotic limits in a 1D quantum walk not with two coins but with a single three-state coin(qutrit). The paper is organized in the following manner, starting with a brief introduction and motivation, in the next section we discuss how Parrondo's games cannot be implemented in asymptotic limits with a 1D qubit quantum walk and then in the following section we discuss 1D quantum walks with a qutrit and an implementation of true Parrondo's games in QWs with qutrits is devised. In the section on Discussions we discuss the reasons behind why a three-state coin(qutrit) delivers a genuine Parrondo's paradox while a two-state coin(qubit) doesn't and finally  we end with a conclusion.  
\\

\begin{figure}
\centering
\includegraphics[scale=0.5]{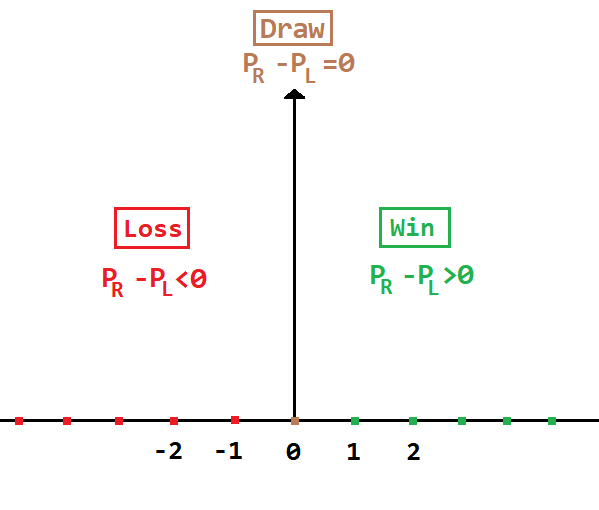}
\caption{An illustration for win or loss conditions for 1-D QW}
\label{fig:win-loss}
\end{figure}

\section{Parrondo's games in quantum walks with a qubit (two state coin)}\label{2state}
In this section, we show why the implementation of Parrondo's paradox failed in case of quantum walks with a qubit as shown in \cite{minli,flitney}. { This part is included as a motivation for our next section to implement a true Parrondo's paradox in 1D quantum walks with a qutrit}.  To understand the reasons behind the failure to see a genuine Parrondo's paradox in quantum walks with a single qubit, we proceed as follows: Two unitary operators $U(\alpha_{A},\beta_{A},\gamma_{A})$ and $U(\alpha_{B},\beta_{B},\gamma_{B})$, representing two games A and B are alternately played in each time step-
\begin{equation}
U(\alpha,\beta,\gamma)=\left(\begin{array}{cc}
e^{i\alpha}\cos\beta & -e^{-i\gamma}\sin\beta\\
e^{i\gamma}\sin\beta & e^{-i\alpha}\cos\beta
\end{array}\right).\label{eq:SU(2)}
\end{equation}
The quantum walker is in an initial state $\vert\Psi_{0}\rangle=\frac{1}{\sqrt{2}}\vert 0\rangle_p \otimes(\vert 0\rangle -i\vert 1\rangle)_{c},$ where subscript $p$ refers to the position space and $c$ refers to the single two state coin (qubit)  space which is initially in  a superposition of $|0\rangle=\left[ \begin{array}{c}
1 \\ 
0
\end{array}\right]$ and $|1\rangle = \left[\begin{array}{c}
0 \\ 
1
\end{array}\right]$. The dynamics of the walker is governed by an unitary shift operator ($\mathcal{S}$) acting in the position space defined as,
\begin{equation}
\mathcal{S} = \!\!\!  \sum\limits_{n=-\infty}^{\infty}\!\!\! \vert n+1 \rangle_p \langle n \vert_p \otimes \vert 0 \rangle \langle 0 \vert + \!\!\!   \sum\limits_{n=-\infty}^{\infty}\!\!\! \vert n-1 \rangle_p \langle n \vert_p \otimes \vert 1 \rangle \langle 1 \vert .
\label{Equ:S}
\end{equation}
The games \emph{A} and \emph{B} can be played in any sequence, i.e., $U(\alpha_{A},\beta_{A},\gamma_{A})$(representing game \emph{A}) is operated on time steps $t=nq$ and $U(\alpha_{B},\beta_{B},\gamma_{B})$(representing game \emph{B}) is operated on time steps $t\neq nq$, where $q$ is the period and $n$ is an integer. The above scheme produces a game sequence- $ABAB\ldots$ for $ q=2$. Similarly, game sequences such as $ABBABB\ldots$ for $q=2$, etc., can also be generated. The evolution operator combining shift and coin operators can be written as: 
\begin{equation}
U =\left\lbrace 
	\begin{array}{ll}
		\mathcal{S}\cdot (I_p \otimes U(\alpha_{A},\beta_{A},\gamma_{A}))  & \mbox{if } t=nq,n\in Z \\
		\mathcal{S}\cdot (I_p \otimes U(\alpha_{B},\beta_{B},\gamma_{B})) & \mbox{if } t\neq nq,n\in Z
	\end{array}
\right. 
\end{equation}
\begin{figure} 
\includegraphics[scale=0.075]{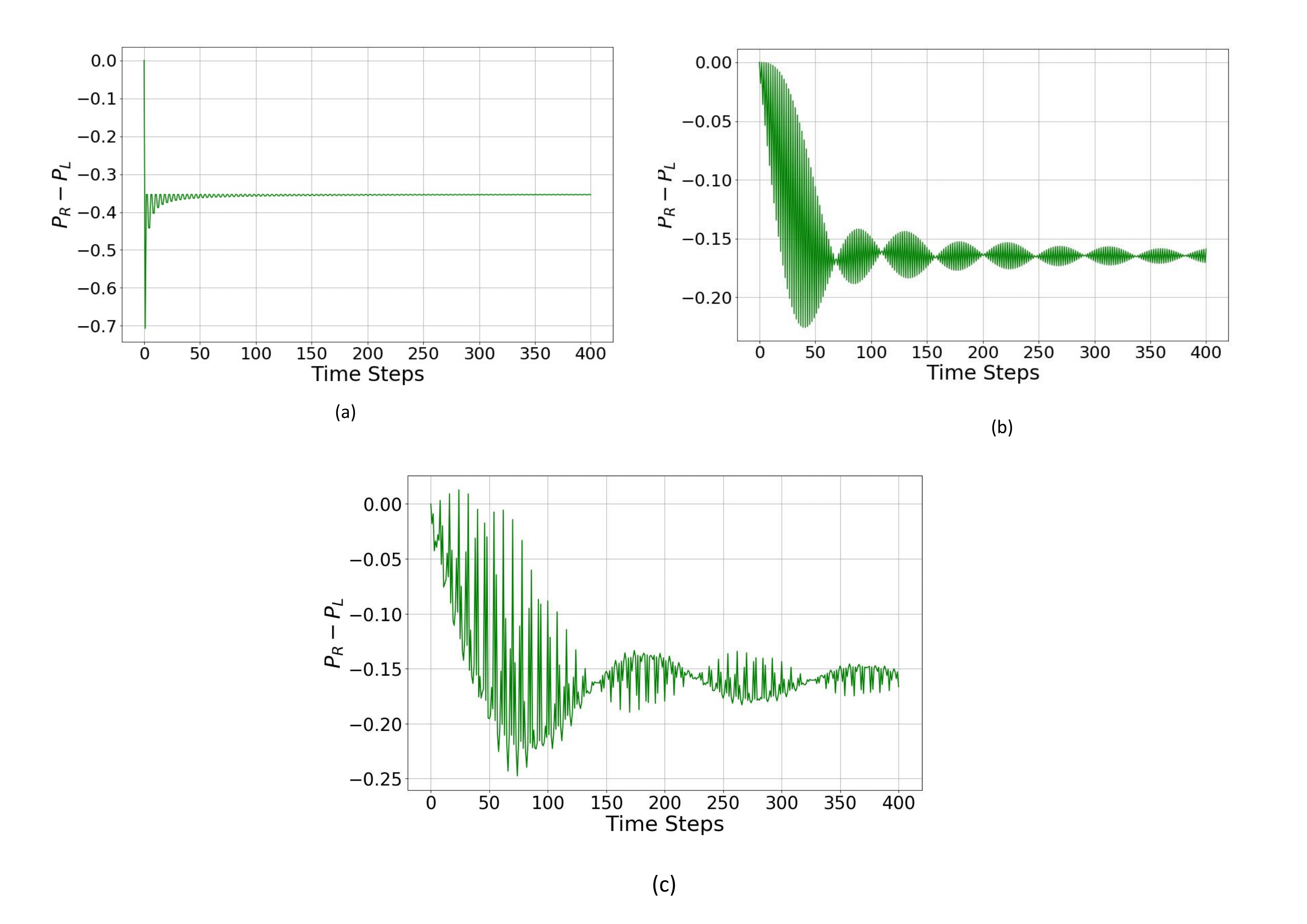}
 \caption{a) $P_R -P_L$ for the sequence $AAAA\ldots$ with initial state $| 0 \rangle_p \otimes \frac{1}{\sqrt{2}}(| 0 \rangle_q - i| 1 \rangle_q)$ with coin operator $A = U(-45, 45, 0)$,  b) $P_R -P_L$ for the sequence $BBBB\ldots$ with initial state $| 0 \rangle_p \otimes \frac{1}{\sqrt{2}}(| 0 \rangle_q - i| 1 \rangle_q)$ with coin operator $B = U(0, 88, -15)$, c) $P_R -P_L$ for the sequence $ABBBABBB\ldots$ with initial state $| 0 \rangle_p \otimes \frac{1}{\sqrt{2}}(| 0 \rangle_q - i| 1 \rangle_q)$ with coin operator $A = U(-45, 40, 0)$ and $B = U(0, 88, -15)$}\label{previous_result}
\end{figure}
and after $N$ steps the final state is- $\vert\Psi_{N}\rangle=U^{N}\vert\Psi_{0}\rangle$. As illustrated in Fig. \ref{fig:win-loss}, after $N$ steps, if the probability of finding the walker to the right of the origin $P_{R}$, is greater than the probability $P_{L}$ to be found to the left of the origin, i.e., $P_{R}-P_{L}>0$, then the player wins. Similarly, if $P_{R}-P_{L}<0$, the player losses. If $P_{R}-P_{L}=0$, it's a draw. Parrondo's games using 1D discrete time QW are formulated by making use of above scheme. Two different coin operators $U_{A} (\alpha_{A},\beta_{A},\gamma_{A})$ and $U_{B}(\alpha_{B},\beta_{B},\gamma_{B})$ are used to construct the two losing games \emph{AAA..} and \emph{BBB..}, for $\alpha_A=-45,\beta_{A}=45,\gamma_{A}=0,$ AND $\alpha_{B}=0,\beta_{B}=88,\gamma_B=-15$, i.e., for $U_{A}=U(-45,45,0)$, $U_{B}=U(0,88,-15)$ we obtain the Fig.\ref{previous_result}. Individually each of the games is losing (See Fig.~\ref{previous_result}(a),(b))and for the sequence- \emph{$ABBBABBB\ldots$} the result is winning at the beginning but in the asymptotic limit the player loses again as in Fig.~\ref{previous_result}(c), one can check for different sequences, like- \emph{$ABAB\ldots, ABBABB\ldots$, etc.,} and in all cases in the asymptotic limits one loses. Hence Parrondo's paradox does not exist in case of 1-D discrete time QW with a single qubit as also established in Refs.~\cite{minli, flitney}. 

To see a Parrondo's paradox in 1D discrete quantum walks is thus the main motivation of this work and we show in the next section a recipe to do just that using a single three-state coin (qutrit). 
\section{True Parrondo's games in quantum walks with a qutrit(three state coin)}\label{3state}
In this section, we define a one-dimensional discrete time QW with a three-state coin. The position of the walker is defined by a vector in the Hilbert space $\mathcal{H_P}$ spanned by an orthogonal normal basis $\lbrace \vert n \rangle : n \in \mathbb{Z} \rbrace$. At each position, the walker is in a superposition of three coin basis states. This coin state is a vector in Hilbert space $\mathcal{H_C}$ with basis states represented by the following orthogonal vectors:
\begin{equation}
|1\rangle= \left[  \begin{array}{c}
1 \\ 
0 \\ 
0
\end{array} \right]  , |0\rangle= \left[  \begin{array}{c}
0 \\ 
1 \\ 
0
\end{array} \right]  , |2\rangle= \left[  \begin{array}{c}
0 \\ 
0 \\ 
1
\end{array} \right]
\end{equation} 
The state of the walker at each time step $t \in \left\lbrace 0,1,2 \ldots \right\rbrace $ $| \psi_t \rangle$ is defined on the Hilbert space $\mathcal{H_P} \otimes \mathcal{H_C}$, the initial state of the walker at time step $t=0$ is 
{
$|\psi_0 \rangle = | 0 \rangle_p \otimes \mathlarger{\mathlarger{\vert \chi \rangle_c}} $}
\begin{equation}\label{initial_1}
|\psi_0 \rangle = | 0 \rangle_p \otimes \frac{1}{\sqrt{3}}(| 0 \rangle + | 1 \rangle - i| 2 \rangle)
\end{equation}
{where, $\vert \chi \rangle_c = \frac{1}{\sqrt{3}}(| 0 \rangle + | 1 \rangle - i| 2 \rangle)$ is the initial coin state.}\\ 
A 1D QW with qutrit is defined with the position of the walker shifted by the shift operator $\mathcal{S^{\prime}}$ after the coin operator $\mathcal{C}$ is operated on the qutrit state as:
{\begin{equation}
\vert\psi_{t+1}\rangle= \mathcal{S^{\prime}} \cdot ( \mathbb{I}\otimes\mathcal{C}) \vert\psi_{t}\rangle
\end{equation}}
where,
\begin{eqnarray}\label{shift_3}
\mathcal{S^{\prime}} \!\!&=&  \!\! \sum\limits_{n=-\infty}^{\infty}\vert n+1 \rangle_p \langle n \vert_p \otimes \vert 0 \rangle \langle 0 \vert \!\!+\!\!   \sum\limits_{n=-\infty}^{\infty}\vert n \rangle_p \langle n \vert_p \otimes \vert 1 \rangle \langle 1 \vert \nonumber\\
&+&   \sum\limits_{n=-\infty}^{\infty}\vert n-1 \rangle_p \langle n \vert_p \otimes \vert 2 \rangle \langle 2 \vert .
\end{eqnarray}
The shift operator is defined in such a way that only when the coin is in the state $| 0 \rangle$ or $| 2\rangle $ the walker moves else when walker is in state $|1\rangle$ the walker stays in the same position. The coin operator $\mathcal{C}$ is defined as follows:
\begin{equation}
\label{coin_oper}
\mathcal{C}=\mathcal{C}(\alpha, \beta, \gamma, \theta)=\left(  \begin{array}{ccc}
I & J & K \\ 
K & I & J \\ 
J & K & I 
\end{array} \right) + i\left(  \begin{array}{ccc}
R & B & G \\ 
B & G & R \\ 
G & R & B 
\end{array} \right)
\end{equation}
where $\alpha$, $\beta$, $\gamma$ and $\theta$ are four real parameters defining the elements: $I, J, K, R, G$ and $B$ with
\begin{equation}
\begin{array}{ll}
3I &= \cos(\gamma) + 2\cos(\theta)\cos(\alpha) \\
3J &= \cos(\gamma) + 2\cos(\theta)\cos(\alpha + 2\pi/3)\\
3K &= \cos(\gamma) + 2\cos(\theta)\cos(\alpha + \pi/3) \\
3R &= \sin(\gamma) + 2\sin(\theta)\cos(\beta) \\
3G &= \sin(\gamma) + 2\sin(\theta)\cos(\beta + 2\pi/3) \\
3B &= \sin(\gamma) + 2\sin(\theta)\cos(\beta + 4\pi/3) 
\end{array}
\end{equation}
This kind of unitary operators are used in particle physics in the density matrix formalism for elementary particles\cite{particle_phy}. Similar to that of single two state coin or qubit QW scheme discussed in preceding  section, we define an evolution operator as follows:
\begin{equation}
U^{\prime}=\left\lbrace 
	\begin{array}{ll}
		\!\!\mathcal{S^{\prime}}\cdot (I_p \otimes \mathcal{C}(\alpha_A, \beta_A, \gamma_A, \theta_A)),  & \!\!\mbox{if } t=nq,n\in Z, \\
		\!\!\mathcal{S^{\prime}}\cdot (I_p \otimes \mathcal{C}(\alpha_B, \beta_B, \gamma_B, \theta_B)),  & \!\!\mbox{if } t\neq nq,n\in Z.
	\end{array}
\right.  
\end{equation}
The evolution of the walker after $N$ steps is $|\psi_N\rangle = U^{\prime N}|\psi_0\rangle$. Similar to that of the two state coin quantum walks, winning and losing is defined as follows: if the probability of finding the walker to the right of the origin $P_{R}$, is greater than the probability $P_{L}$ to be found to the left of the origin , i.e., $P_{R}-P_{L}>0$, then the player wins. Similarly, if $P_{R}-P_{L}<0$, the player losses. If $P_{R}-P_{L}=0$, it's a draw. The two coin operators corresponding to the two games $A$ and $B$ are defined as follows:
\begin{equation}\label{coin_operator}
\begin{array}{ll}
A &= \mathcal{C}(\alpha_A, \beta_A, \gamma_A, \theta_A) = \mathcal{C}(\pi, \frac{\pi}{2}, \pi, \pi) , \\
B &= \mathcal{C}(\alpha_B, \beta_B, \gamma_B, \theta_B) = \mathcal{C}(\frac{\pi}{2}, \frac{\pi}{2}, \frac{3\pi}{2}, \frac{\pi}{2}).
\end{array}
\end{equation}
\begin{figure} 
\includegraphics[scale=0.3]{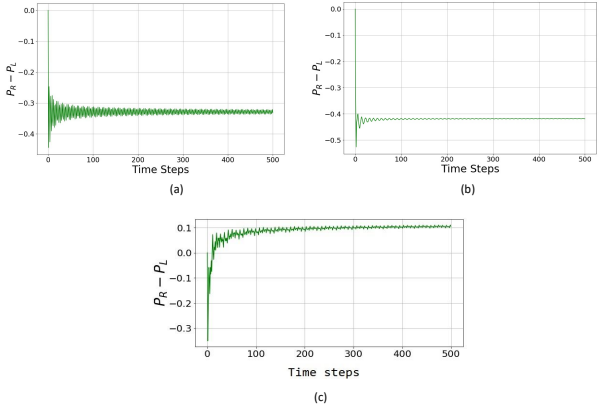}
 \caption{a) $P_R -P_L$ for the sequence $AAAA\ldots$ with initial state $| 0 \rangle_p \otimes \frac{1}{\sqrt{3}}(| 0 \rangle + | 1 \rangle - i| 2 \rangle)$ with coin operator $A = \mathcal{C}(\pi, \pi/2, \pi, \pi)$  b) $P_R -P_L$ for the sequence $BBBB\ldots$ with initial state $| 0 \rangle_p \otimes \frac{1}{\sqrt{3}}(| 0 \rangle + | 1 \rangle - i| 2 \rangle)$ with coin operator $B = \mathcal{C}(\pi/2, \pi/2, 3\pi/2, \pi/2)$ c) $P_R -P_L$ for the sequence $ABAB\ldots$ with initial state $| 0 \rangle_p \otimes \frac{1}{\sqrt{3}}(| 0 \rangle + | 1 \rangle - i| 2 \rangle)$ with coin operator $A = \mathcal{C}(\pi, \pi/2, \pi, \pi)$ and $B = \mathcal{C}(\pi/2, \pi/2, 3\pi/2, \pi/2)$ for 500 steps}\label{3state_result}
\end{figure}
Here $A$ is operated on time steps $t=nq$ and $B$ is played on time steps $t\neq nq$, where $q$ is the period and $n$ is an integer same as discussed in preceding section. For the choice of $\alpha_A=\pi,\beta_{A}=\pi/2,\gamma_{A}=\pi,\theta_{A}=\pi, \alpha_{B}=\pi/2,\beta_{B}=\pi/2,\gamma_B=3\pi/2, \theta=\pi/2$ as shown in Eq.~\ref{coin_operator} we obtain Fig.~\ref{3state_result}. When game AAAA... and BBBB... are played it results in losing (see Figs.~\ref{3state_result}(a),(b)), whereas when they were played in the sequence $ABAB\ldots$ we obtain a winning outcome (see Fig.~\ref{3state_result}(c)). Thus unlike a two state coin(qubit), in case of a three state coin(qutrit) in asymptotic limits we obtain a true Parrondo's paradox.
\section{Discussion}\label{discussion}
When a two-state coin(qubit) was considered the Parrondo's games did not give rise to the paradox in asymptotic limits of the 1D QW (see Fig.~\ref{previous_result}) whereas when a three-state coin is used for quantum walks we obtain a true Parrondo's paradox. In order to obtain a true Parrondo's paradox, a three-state coin is needed. In Refs.\cite{minli,flitney} it was shown that in asymptotic limits for a two-state coin the paradox does not exist. To identify the reasons for the success of three state coin as compared to a two-state coin we first study the influence of the initial state in Parrondo's games. Let us consider a different initial state given as:
\begin{equation}\label{new_initial}
|\psi_0 \rangle = | 0 \rangle_p \otimes \frac{1}{\sqrt{3}}(| 1 \rangle +| 0 \rangle - | 2 \rangle)
\end{equation} 
\begin{figure} 
\includegraphics[scale=0.3]{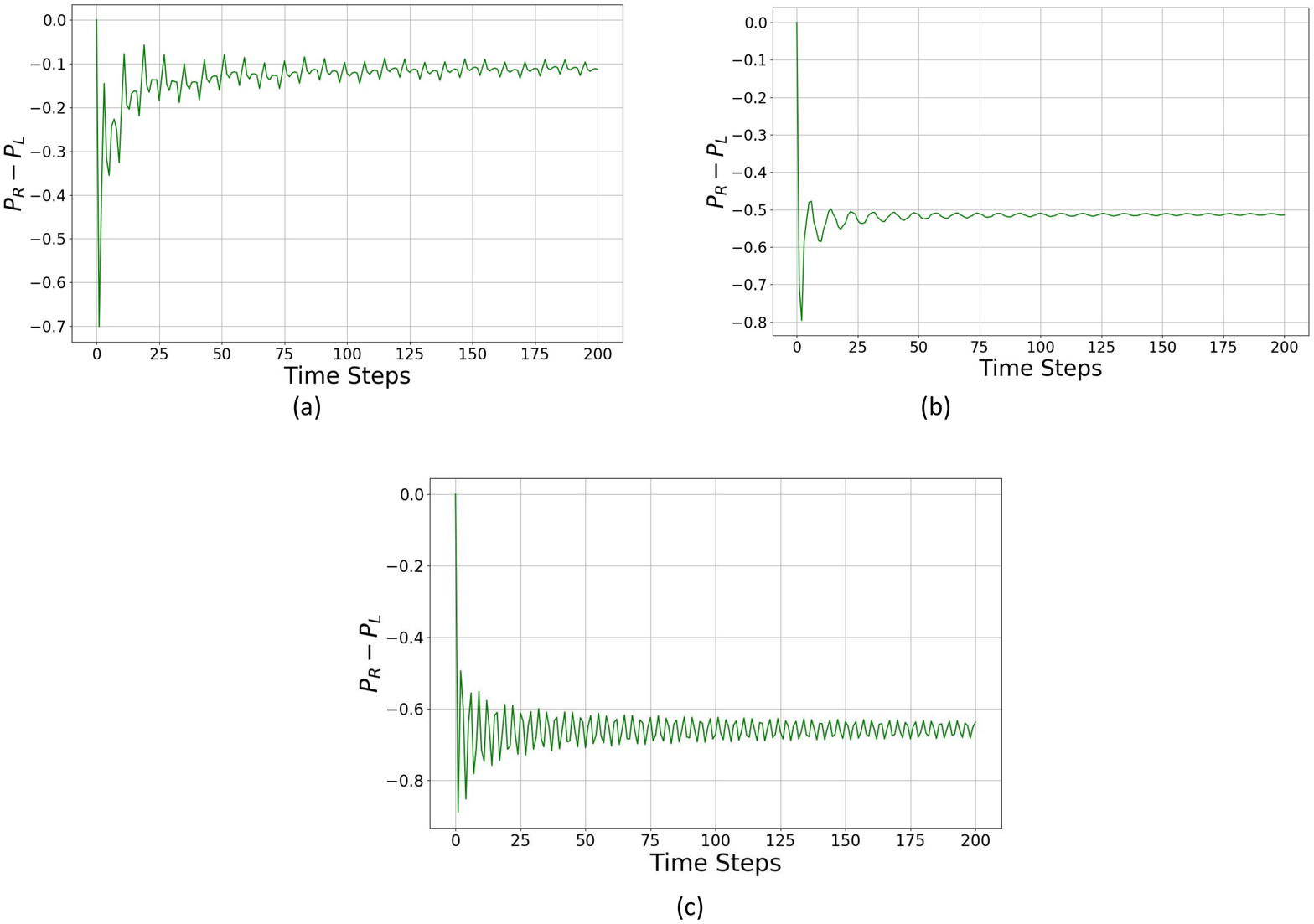}
 \caption{a) $P_R -P_L$ for the sequence $AAAA\ldots$ with initial state $| 0 \rangle_p \otimes \frac{1}{\sqrt{3}}(| 1 \rangle + | 0 \rangle - | 2 \rangle)$ with coin operator $A = \mathcal{C}(\pi, \pi/2, \pi, \pi)$  b) $P_R -P_L$ for the sequence $BBBB\ldots$ with initial state $| 0 \rangle_p \otimes \frac{1}{\sqrt{3}}(| 1 \rangle + | 0 \rangle - | 2 \rangle)$ with coin operator $B = \mathcal{C}(\pi/2, \pi/2, 3\pi/2, \pi/2)$ c) $P_R -P_L$ for the sequence $ABAB\ldots$ with initial state $| 0 \rangle_p \otimes \frac{1}{\sqrt{3}}(| 1 \rangle + | 0 \rangle - | 2 \rangle)$ with coin operator $A = \mathcal{C}(\pi, \pi/2, \pi, \pi)$ and $B = \mathcal{C}(\pi/2, \pi/2, 3\pi/2, \pi/2)$ for 200 steps}\label{initial_state}
\end{figure}
which is different from the initial state (Eq.~\ref{initial_1}) considered in the preceding section. Now with this new initial state but with the same shift $\mathcal{S^{\prime}}$ (Eq.~\ref{shift_3}), we now obtain Fig.~\ref{initial_state}. It is clear that the new initial state does not give us a paradox, when the games are played in sequence $ABAB\ldots$(see Fig.~\ref{initial_state}(c)). {Now considering again another initial state given as:
\begin{equation}\label{new_initial1}
|\psi_0 \rangle = | 0 \rangle_p \otimes \frac{1}{\sqrt{3}}(i| 1 \rangle +| 0 \rangle - | 2 \rangle)
\end{equation} 
\begin{figure} 
 \includegraphics[scale=0.075]{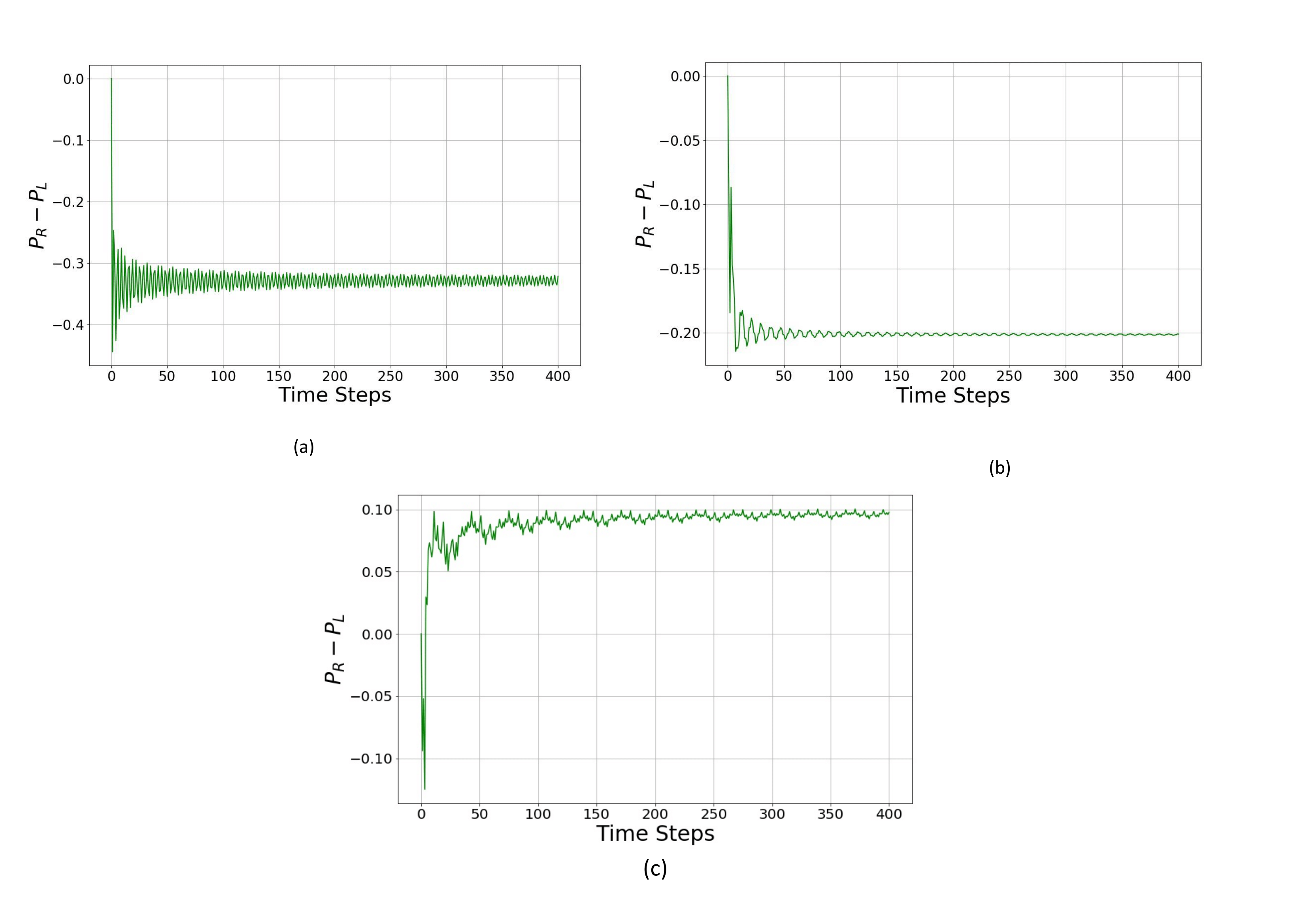}
 \caption{{a) $P_R -P_L$ for the sequence $AAAA\ldots$ with initial state $| 0 \rangle_p \otimes \frac{1}{\sqrt{3}}(i| 0 \rangle + | 1 \rangle - | 2 \rangle)$ with coin operator $A = \mathcal{C}(\pi, \pi/2, \pi, \pi)$  b) $P_R -P_L$ for the sequence $BBBB\ldots$ with initial state $| 0 \rangle_p \otimes \frac{1}{\sqrt{3}}(i| 0 \rangle + | 1 \rangle - | 2 \rangle)$ with coin operator $B = \mathcal{C}(\pi/2, \pi/2, 3\pi/2, \pi/2)$ c) $P_R -P_L$ for the sequence $ABAB\ldots$ with initial state $| 0 \rangle_p \otimes \frac{1}{\sqrt{3}}(i| 0 \rangle + | 1 \rangle - | 2 \rangle)$ with coin operator $A = \mathcal{C}(\pi, \pi/2, \pi, \pi)$ and $B = \mathcal{C}(\pi/2, \pi/2, 3\pi/2, \pi/2)$ for 400 steps}}\label{i11}
\end{figure}
in this case, plot of the median $P_R-P_L$ shown in Fig.~\ref{i11} results in a Parrondo's paradox.} This implies that the initial state plays a crucial role in the Parrondo's games, with different initial states one may or may not obtain a true Parrondo's paradox.

Next we consider the influence of the shift operator, using a different shift operator for the same initial state as in Eq.~(\ref{initial_1}). Here we define a new shift $\mathcal{S^{\prime}_\texttt{1}}$-
\begin{eqnarray}\label{new_shift}
\mathcal{S^{\prime}_\texttt{1}} \!\!&=& \!\!  \sum\limits_{n=-\infty}^{\infty}\vert n+1 \rangle_p \langle n \vert_p \otimes \vert  1\rangle \langle 1 \vert +   \sum\limits_{n=-\infty}^{\infty}\vert n \rangle_p \langle n \vert_p \otimes \vert 0 \rangle \langle 0 \vert \nonumber\\
&+&   \sum\limits_{n=-\infty}^{\infty}\vert n-1 \rangle_p \langle n \vert_p \otimes \vert 2 \rangle \langle 2 \vert
\end{eqnarray}
\begin{figure} 
\includegraphics[scale=0.3]{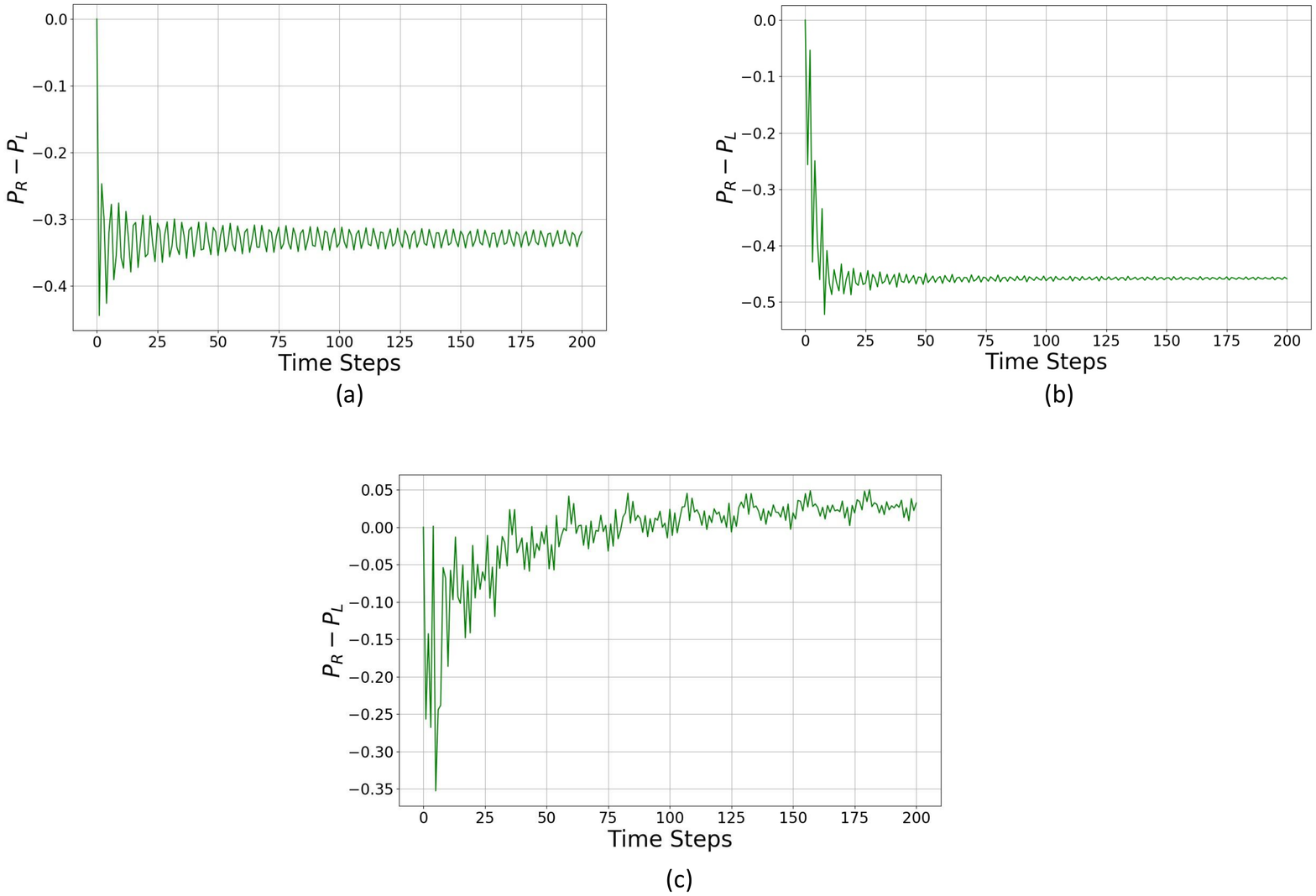}
 \caption{a) $P_R -P_L$ for the sequence $AAAA\ldots$ with initial state $| 0 \rangle_p \otimes \frac{1}{\sqrt{3}}(| 1 \rangle + | 0 \rangle - i | 2 \rangle)$ with coin operator $A = \mathcal{C}(\pi, \pi/2, \pi, \pi)$  b) $P_R -P_L$ for the sequence $BBBB\ldots$ with initial state $| 0 \rangle_p \otimes \frac{1}{\sqrt{3}}(| 1 \rangle + | 0 \rangle - i| 2 \rangle)$ with coin operator $B = \mathcal{C}(\pi/2, \pi/2, 3\pi/2, \pi/2)$ c) $P_R -P_L$ for the sequence $ABAB\ldots$ with initial state $| 0 \rangle_p \otimes \frac{1}{\sqrt{3}}(| 1 \rangle + | 0 \rangle - i| 2 \rangle)$ with coin operator $A = \mathcal{C}(\pi, \pi/2, \pi, \pi)$ and $B = \mathcal{C}(\pi/2, \pi/2, 3\pi/2, \pi/2)$ for the shift operator $\mathcal{S^{\prime}_\texttt{1}}$ defined in Eq. \ref{new_shift}}\label{shift1}
\end{figure}
The shift operator as defined in Eq.~\ref{new_shift} also gives us a Parrondo's paradox after around $200$ steps(almost asymptotic limit) as in Fig.~\ref{shift1}. One can see that the change in shift operator has changed the behavior of the Parrondo's games, here the yield of the player is changing. Now if we consider another shift operator $\mathcal{S^{\prime}_\texttt{2}}$ 
\begin{eqnarray}\label{new_shift2}
\mathcal{S^{\prime}_\texttt{2}}\!\!\!&=&\!\!\!\!\!\!\sum\limits_{n=-\infty}^{\infty}\!\!\!\!\vert n+1 \rangle_p \langle n \vert_p \otimes \vert  1\rangle \langle 1 \vert \!\!+\!\!\!\!\!\!\sum\limits_{n=-\infty}^{\infty}\!\!\!\vert n-1 \rangle_p \langle n \vert_p \otimes \vert 0 \rangle \langle 0 \vert \nonumber\\
&+&  \!\!\! \sum\limits_{n=-\infty}^{\infty}\vert n \rangle_p \langle n \vert_p \otimes \vert 2 \rangle \langle 2 \vert
\end{eqnarray}
\begin{figure} 
\includegraphics[scale=0.3]{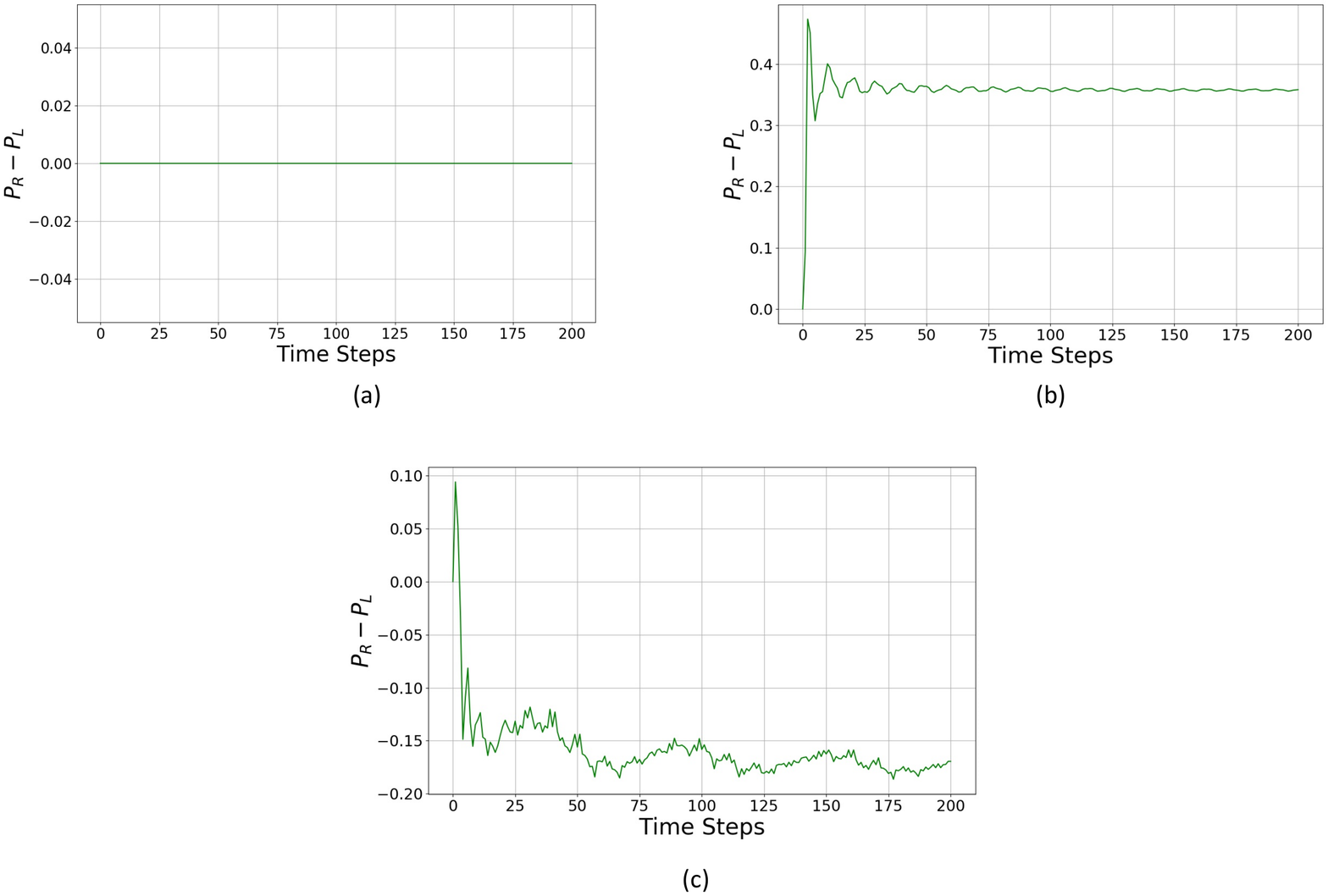}
\caption{a) $P_R -P_L$ for the sequence $AAAA\ldots$ with initial state $| 0 \rangle_p \otimes \frac{1}{\sqrt{3}}(| 1 \rangle + | 0 \rangle - i | 2 \rangle)$ with coin operator $A = \mathcal{C}(\pi, \pi/2, \pi, \pi)$  b) $P_R -P_L$ for the sequence $BBBB\ldots$ with initial state $| 0 \rangle_p \otimes \frac{1}{\sqrt{3}}(| 1 \rangle + | 0 \rangle - i| 2 \rangle)$ with coin operator $B = \mathcal{C}(\pi/2, \pi/2, 3\pi/2, \pi/2)$ c) $P_R -P_L$ for the sequence $ABAB\ldots$ with initial state $| 0 \rangle_p \otimes \frac{1}{\sqrt{3}}(| 1 \rangle + | 0 \rangle - i| 2 \rangle)$ with coin operator $A = \mathcal{C}(\pi, \pi/2, \pi, \pi)$ and $B = \mathcal{C}(\pi/2, \pi/2, 3\pi/2, \pi/2)$ for the shift operator $\mathcal{S^{\prime}_\texttt{1}}$ defined in Eq. \ref{new_shift2}}\label{shift2}
\end{figure}
where the state $\vert 2 \rangle$ is the wait state, we obtain the plot of median, $P_R-P_L$ as in Fig.~\ref{shift2}. Here the game sequence $AAAA\ldots$ gives us a draw and the sequence $BBBB\ldots$ provides a winning outcome whereas the $ABAB\ldots$ sequence gives us a losing outcome, which is also a paradox with a role reversal with the definitions of win and loss from Fig.~\ref{fig:win-loss} is reversed. Thus we can conclude that the change of shift operator does not affect the paradox but only the yield of the player. { So what does the qutrit QW  have which the qubit QW doesn't have? To understand this we focus on the difference between the shift operators used for qubit versus that used for qutrit case.  In case of a qutrit, there are 3 states, and the shift operators: $\mathcal{S^\prime}, \mathcal{S}_{1}^\prime, \mathcal{S}_{2}^\prime$ is of the form as given in Eqs.~(\ref{shift_3},\ref{new_shift},\ref{new_shift2}), have a wait state, a left going and another right going state. In qubit quantum walk, the shift operator is defined as in Eq.~\ref{Equ:S}, $\mathcal{S}$ is bereft of any wait state. Thus, the most plausible reason behind observing the paradox in qutrits but not in qubits is the possibility of a wait state. Of course the initial state of qutrit with three coin states as opposed to only two possible coin states for a qubit has an important bearing in observing Parrondo's paradox as shown in Figs.~{\ref{3state_result},\ref{initial_state}}}.
\begin{figure} 
 \includegraphics[scale=0.075]{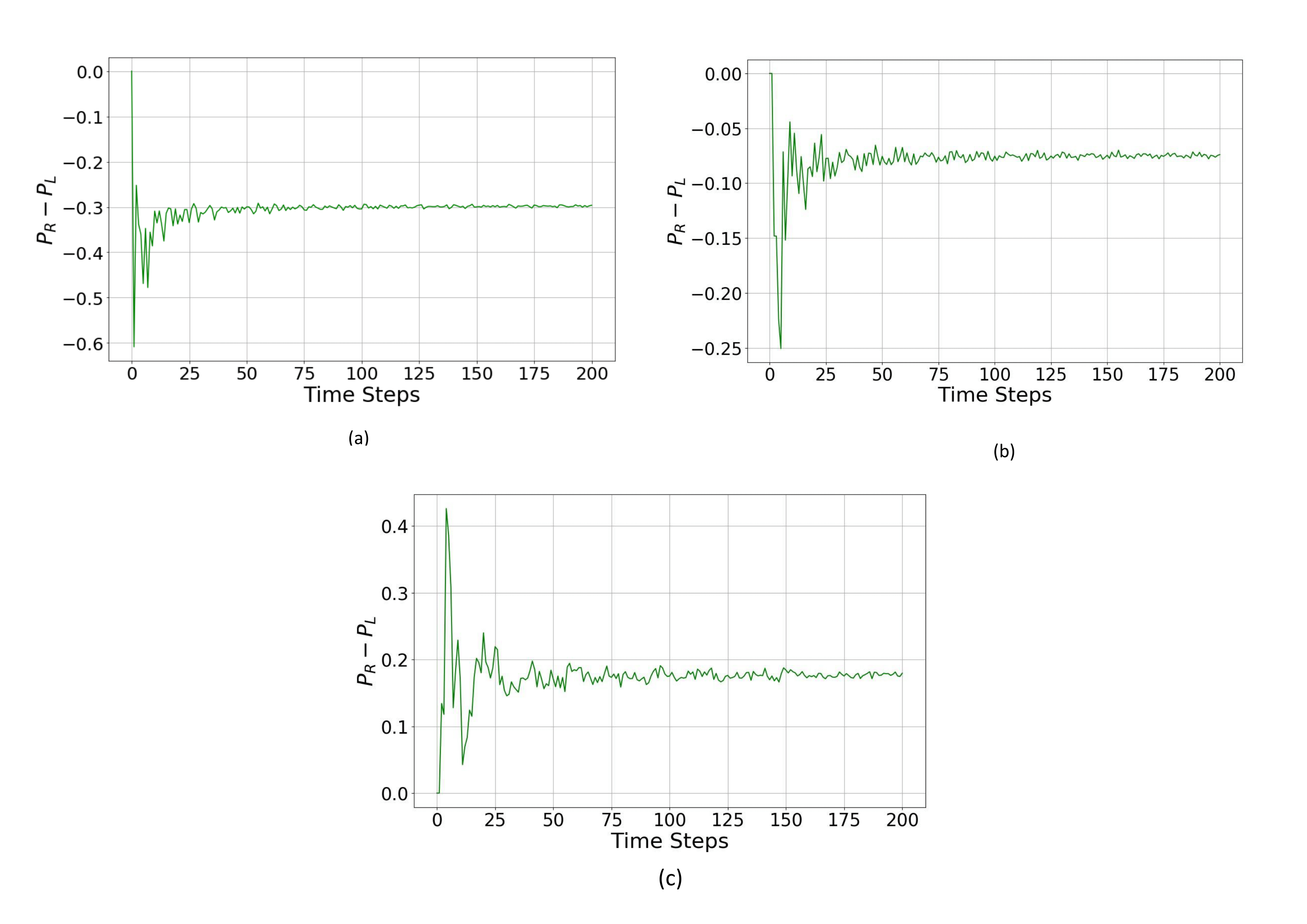}
 \caption{
 {(a) $P_L$ for the sequence $AAAA\ldots$ with initial state $| 0 \rangle_p \otimes \frac{1}{\sqrt{3}}(| 0 \rangle + | 1 \rangle - i| 2 \rangle)$ with coin operator $A = \mathcal{C}(\dfrac{\pi}{8}, \dfrac{3\pi}{8}, \dfrac{3\pi}{4}, \dfrac{\pi}{4})$,  (b) $P_R -P_L$ for the sequence $BBBB\ldots$ with initial state $| 0 \rangle_p \otimes \frac{1}{\sqrt{3}}(| 0 \rangle + | 1 \rangle - i| 2 \rangle)$ with coin operator $B = \mathcal{C}(\dfrac{2\pi}{3}, 7\pi, \dfrac{3\pi}{2}, 2\pi)$, (c) $P_R -P_L$ for the sequence $ABAB\ldots$ with initial state $| 0 \rangle_p \otimes \frac{1}{\sqrt{3}}(| 0 \rangle + | 1 \rangle - i| 2 \rangle)$ with coin operator $A = \mathcal{C}(\dfrac{\pi}{8}, \dfrac{3\pi}{8}, \dfrac{3\pi}{4}, \dfrac{\pi}{4})$ and $B = \mathcal{C}(\dfrac{2\pi}{3}, 7\pi, \dfrac{3\pi}{2}, 2\pi)$ for 400 steps for shift operator $\mathcal{S^{\prime}}$ as defined in Eq.~\ref{shift_3}}}\label{anglework}
\end{figure}

\begin{figure} 
 \includegraphics[scale=0.075]{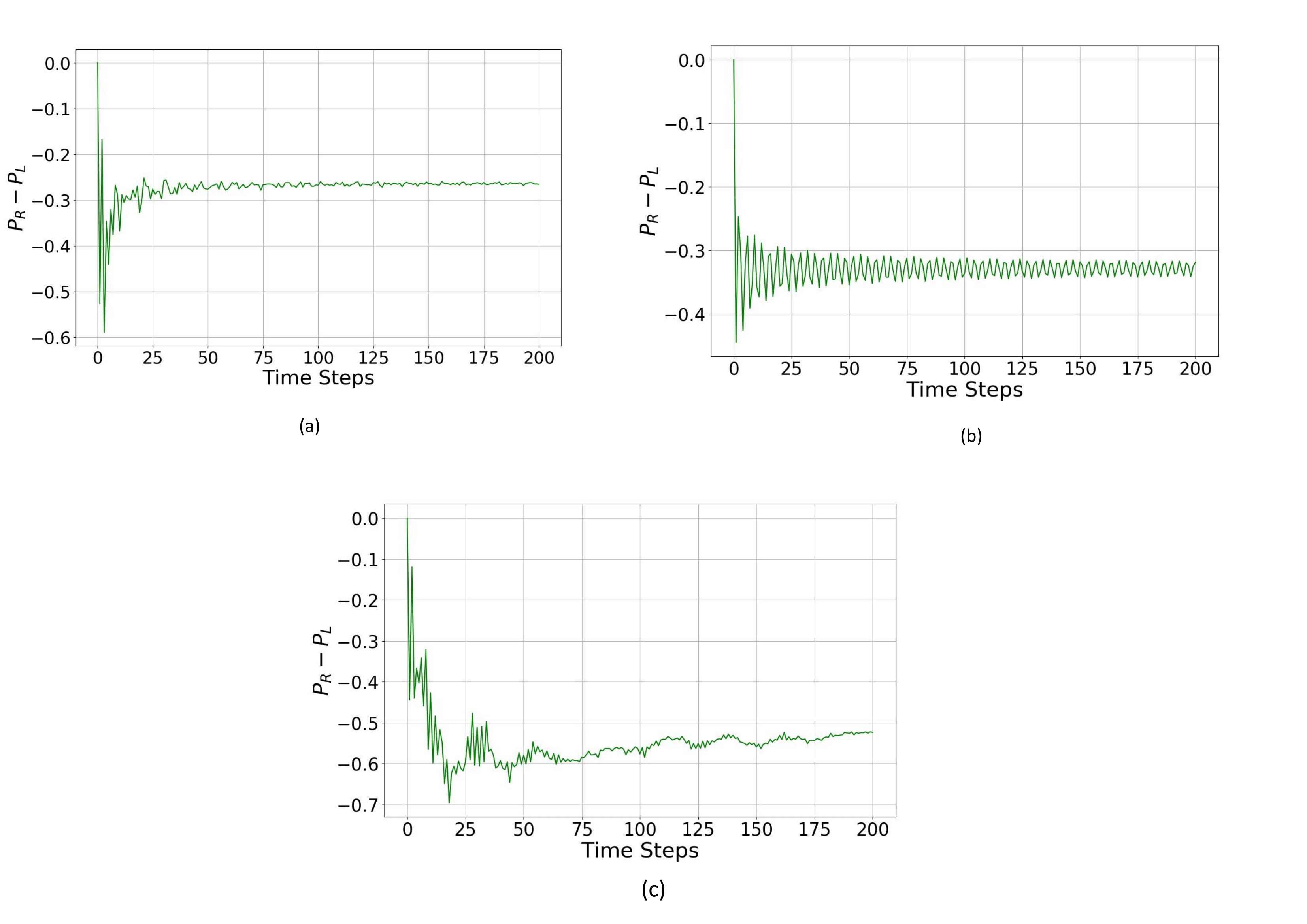}
 \caption{{{a) $P_R -P_L$ for the sequence $AAAA\ldots$ with initial state $| 0 \rangle_p \otimes \frac{1}{\sqrt{3}}(| 0 \rangle + | 1 \rangle - i| 2 \rangle)$ with coin operator $A = \mathcal{C}(\dfrac{\pi}{8}, \dfrac{2\pi}{8}, \dfrac{3\pi}{4}, \dfrac{6\pi}{4})$, (b) $P_R -P_L$ for the sequence $BBBB\ldots$ with initial state $| 0 \rangle_p \otimes \frac{1}{\sqrt{3}}(| 0 \rangle + | 1 \rangle - i| 2 \rangle)$ with coin operator $B = \mathcal{C}(2\pi, 3\pi, 2\pi, \pi)$, (c) $P_R -P_L$ for the sequence $ABAB\ldots$ with initial state $| 0 \rangle_p \otimes \frac{1}{\sqrt{3}}(| 0 \rangle + | 1 \rangle - i| 2 \rangle)$ with coin operator $A = \mathcal{C}(\dfrac{\pi}{8}, \dfrac{2\pi}{8}, \dfrac{3\pi}{4}, \dfrac{6\pi}{4})$ and $B = \mathcal{C}(2\pi, 3\pi, 2\pi, \pi)$ for 400 steps for shift operator $\mathcal{S^{\prime}}$ as defined in Eq. \ref{shift_3}}}}\label{angle_not_working}
\end{figure}

{Finally, we consider the role of the qutrit coin operators as defined in Eq.~(\ref{coin_operator}). In Figs.~\ref{anglework},\ref{angle_not_working}, we can see that for different parameters as in Eq.~(\ref{coin_oper}) we can change the outcome of the game, i.e., we may or may not obtain a Parrondo's paradox. Fig.~8 shows Parrondo's paradox for different qutrit coin operators $A=\mathcal{C}(\dfrac{\pi}{8}, \dfrac{3\pi}{8}, \dfrac{3\pi}{4}, \dfrac{\pi}{4})$ and $B=\mathcal{C}(\dfrac{2\pi}{3}, 7\pi, \dfrac{3\pi}{2}, 2\pi)$. On the other hand Fig.~9 doesn't show Parrondo's paradox for a different set of parameters for the qutrit coin operator $A=\mathcal{C}(\dfrac{\pi}{8}, \dfrac{2\pi}{8}, \dfrac{3\pi}{4}, \dfrac{6\pi}{4})$ and $B=\mathcal{C}(2\pi, 3\pi, 2\pi, \pi)$ as in Eqs.~(8,9). The set of parameters for which we obtain the paradox are not special in any way and one can obtain many such set of parameters for which Parrondo's paradox can be seen.}

{To understand the physical mechanism, one can allude to Ref.\cite{cold_atom1} which shows a quantum walk modeled with cold atoms. In case of cold atoms the shift operators and coin operators are modeled using optical lattice and lasers, such a model can be used to physically realize the shift operators in Eqs.(\ref{shift_3},\ref{new_shift},\ref{new_shift2}). In Ref.\cite{minli} too, it has been shown that Parrondo's paradox does not exist for a single qubit coin quantum walk considered. Qutrits can be physically realized with biphotons, demonstrated in Ref.\cite{biphotonic}, using two correlated photons- a "Biphotonic qutrit". Such a biphotonic qutrit and optical lattice can be used to physically realize a three state quantum walk. Quantum walks with qutrits can be used for quantum algorithms and its non-trivial nature with Parrondo's games can be helpful for better insights into quantum ratchets and quantum algorithms.}

\section{Conclusion}\label{conclusion}
Earlier attempts using a single two-state coin failed in asymptotic limit \cite{minli,flitney}. Here with the aid of a qutrit we successfully implemented a true Parrondo's paradox (see Figs.~ \ref{3state_result},\ref{i11},\ref{anglework}). Quantum Parrondo's games play an important role in quantum ratchets, providing a mechanism for a particle to transport against an applied bias, quantum analogue for Brownian ratchets. Different kinds of quantum walks and its applications can help the community in better understanding and for developing new quantum algorithms.
\acknowledgments
This work was supported by the grant ``Non-local correlations in nanoscale systems: Role of decoherence, interactions, disorder and pairing symmetry'' from SCIENCE \& ENGINEERING RESEARCH BOARD, New Delhi, Government of India, Grant No.  EMR/20l5/001836, Principal Investigator: Dr. Colin Benjamin, National Institute of Science Education and Research, Bhubaneswar, India.


\begin{thebibliography}{99}
\bibitem{two_particle}
A.~Schreiber, et. al., A 2D quantum
walk simulation of two-particle dynamics, Science \textbf{336}, \textbf{55}
(2012).

\bibitem{Marquez-Martin2016}
I.~M\'arquez-Martín, G.~Di Molfetta, and A.~P\'erez, Fermion confinement via quantum walks in $(2+1)$-dimensional and $(3+1)$-dimensional space-time, \pra \textbf{95}, 042112 (2017)

\bibitem{algo}
A.~Ambainis, Quantum walks and their algorithmic applications, Int. Journal of Quantum Information (\textbf{4}) (2003) \textbf{507}

\bibitem{qcomp1}
A.~M.~Childs, et. al. in: Proceedings of the 35th ACM Symposium on Theory of Computing, ACM Press, New York, 2003, p. \textbf{59}.

\bibitem{qcomp2}
N. Shenvi, J. Kempe, K.~B. Whaley, Quantum random-walk search algorithm. \pra \textbf{67} (2003) \textbf{052307}.

\bibitem{universal}
A.~M.~Childs, Universal Computation by Quantum Walk, Phys. Rev. Lett. \textbf{102}, {180501} (2009). 

\bibitem{classical}
R.~Portugal, Quantum Walks and Search Algorithms, Springer, Berlin 10.1007/978-1-4614-6336-8 (2013) 


\bibitem{Andraca}
S.~E.~Venegas-Andraca, Quantum walks: a comprehensive review , Quantum Information Processing October 2012, Volume \textbf{11}, Issue \textbf{5}, pp \textbf{1015-1106}.

\bibitem{Inui}
N.~Inui, N.~Konno, and E.~Segawa, One-dimensional three-state quantum walk. Phys. Rev. E \textbf{72}, 056112 

\bibitem{qutrit1}
T.~Machida, C.~M.~ Chandrashekar. Localization and limit laws of a three-state alternate quantum walk on a two-dimensional lattice, \pra \textbf{92}, 062307 (2015)

\bibitem{Falkner}
S.~Falkner and S.~Boettcher, Weak limit of the three-state quantum walk on the line, \pra \textbf{90}, 012307 (2014)

\bibitem{qutrit2}
M. \v Stefa\v n\'ak,I. Bezd\v ekov\'a, I. Jex, Limit distributions of three-state quantum walks: the role of coin eigenstates, \pra \textbf{90}, 012342 (2014)

\bibitem{q_game}
David A. Meyer, Quantum Strategies, \prl \textbf{82} (1999) 1052–1055

\bibitem{parrondo}
J.~M.~R.~Parrondo and L.~Dinis, Brownian motion and gambling: from ratchets to paradoxical games, Contemporary Physics,  \textbf{45}, 147 (2004).

\bibitem{maximal_parrondo}
{
F.~A.~Gr\"{u}nbaum, M.~Pejic, Lett Math Phys (2016), \textbf{106}, 251 .}

\bibitem{superactivation}
{
S.~Strelchuk, Parrondo's paradox and superactivation of classical and quantum capacity of communication channels with memory, Phys. Rev. A \textbf{88}, 032311
}

\bibitem{minli}
M.~Li, Y.~S.~Zhang, G.-C.~Guo, Quantum Parrondo's games constructed by quantum random walk, Fluct. Noise Lett. \textbf{12}, \textbf{1350024} (2013).

\bibitem{flitney} 
A.~P.~Flitney, Quantum Parrondo's games using quantum walks, arXiv:\textbf{1209.2252} (2012).

\bibitem{previous}
J. Rajendran and C. Benjamin, Implementing Parrondo's paradox with two coin quantum walks, R. Soc. open sci. 2018 \textbf{5} \textbf{171599}

\bibitem{particle_phy}
C. Brannen, Density Operator Theory and Elementary Particles, available at:\url{http://www.brannenworks.com/densitytime.pdf}

\bibitem{cold_atom1}
S.~Mugel, et. al., Topological bound states of a quantum walk with cold atoms, Phys. Rev. A 94, 023631 (2016)

\bibitem{biphotonic}
B.~P.~Lanyon, et. al., Manipulating Biphotonic Qutrits, Phys. Rev. Lett. \textbf{100}, 060504 (2008)

\end{thebibliography}
\end{document}